\documentstyle[12pt,a4,psfig,fullpage]{article}
\newcommand{\be}{\begin{equation}}
\newcommand{\ee}{\end{equation}}
\newcommand{\beq}{\begin{eqnarray}}
\newcommand{\eeq}{\end{eqnarray}}

\begin{document}
\pagestyle{empty}
\begin{flushright}
{BROWN-HET-1127} \\
{ULG-PNT-98-JRC-3} \\
\end{flushright}
\vspace*{5mm}

\begin{center}
{\Large{\bf Soft Pomeron and Lower-Trajectory Intercepts}}\\
[10mm]

J.R. Cudell$^a$\footnote{JR.Cudell@ulg.ac.be}, V.V.
Ezhela$^b$\footnote{ezhela@mx.ihep.su},
 K. Kang$^c$\footnote{kang@het.brown.edu;
Supported in part by DOE Grant DE-FG02-91ER40688 - Task
A}, S.B. Lugovsky$^b$\footnote{lugovsky@mx.ihep.su},
 and N.P. Tkachenko$^{b}$\footnote{tkachenkon@mx.ihep.su}\\
[5mm]
{\em a.  Inst. de Physique, Universit\'e de Li\`ege, B$\hat{a}$t. B-5,
Sart Tilman, \\B4000 Li\`ege, Belgium} \\
[5mm]
{\em b.  COMPAS Group\footnote{COMPAS Group supported
 in part by RFBR Grant 96-07-89230}, IHEP, Protvino, Russia}\\
[5mm]
{\em c.  Department of Physics, Brown University, Providence RI 02912 USA}\\
[10mm]
(Presented by {\bf Kyungsik Kang}\footnote{at a 4th Workshop on Quantum Chromodynamics, June 1 - 6, 1998, The American University of Paris, Paris, France, and at the 4th Workshop on Small-x and Diffractive Physics, September 17 - 20, 1998, Fermi National Accelerator Laboratory, Batavia, IL})\\
[5mm]
Abstract

\end{center}
\begin{quote}
We present a preliminary report on the determination of the intercepts
and couplings of the soft pomeron and of the $\rho/\omega$ and $f/a$
trajectories from the largest data set available for
all total cross sections and real parts of the hadronic amplitudes.\footnote{A
short preliminary version of this work with some variation has been presented
by
three of us (VVE, SBL and NPT) in the 1998 Review of Particle Physics
\cite{PDG}.} Factorization is reasonably
satisfied by the pomeron couplings, which allows us to make predictions on
$\gamma\gamma$ and $\gamma p$ total cross sections. 
In addition we show that these data cannot discriminate between fits
based on a simple Regge pomeron-pole and on an asymptotic $\log^2s$-type
behaviour, implying that the effect of unitarisation is negligible. Also we
examine the range of validity in energy of the fit, and the bounds that these
data place on the odderon and on the hard pomeron.

\end{quote}
\vspace*{3mm}

\newpage
\setcounter{page}{1}
\pagestyle{plain}
\section{Introduction}
Irrespectively of
the true nature of the pomeron, the Regge parametrisation \cite{DL, Collins}
 plays an important role in the experimental analysis
of diffractive processes at HERA \cite{one} and in $\bar pp$ studies at the
Tevatron and at CERN \cite{two}. It also
offers a successful phenomenological starting point at low $\,
x\,$ and low $\, Q^2\,$ or at the soft limit of the hadronic interactions, 
from
which QCD evolution as well as the soft process theory based on QCD can
in principle be developed  or to which one would have to add
a hard (leading-twist) QCD  contribution \cite{twopoms}. 

Remarkably, the experience over the past several decades in
phenomenological analysis of experiments has shown that the pomeron is, to a
good approximation, a simple Regge pole with intercept $1+\epsilon$; 
despite the apparent complications of
being non-perturbative in QCD, $t=0$ data can be described by particularly
simple models-- namely a sum of simple powers of the center-of-mass energy
$\sqrt{s}$: 
\be
{\it Im A}_{h_1h_2}(s,t)=\sum_i (-1)^{S_i} C_{h_1h_2}(t)
\left({s\over s_0}\right)^{\alpha_i(t)}
\ee
with $S_i$ the signature of the exchange. The total cross section
is then given by
\be
 \sigma_{tot}^{h_1h_2} (s) = {\it Im A}_{h_1h_2}(s,0)/s.
\ee

The trajectories $\alpha_i(t)$ are universal, and the process dependence is
present only in the constants $C_{h_1h_2}$ (which absorb the scale $s_0$).
The trajectories are manifest (and approximately linear) in the case of mesons,
but can only be assumed in the case of the pomeron (despite the existence of
strong glueball candidates \cite{glueballs}).

Adopting the viewpoint that simple Regge pole exchanges should account for all
soft data up to the presently accessible energies, Donnachie and Landshoff (DL)
\cite{four} advanced a model for the total cross sections with just two
Reggeons, i,e., an additional exchange-degenerate Reggeon representing both 
$C=\pm 1$\,   ($\rho$, $\omega$, a, f )\, exchanges besides the pomeron:
\beq
 \sigma_{tot} (s) = X s^{\epsilon} + Y s^{- \eta }
\label{DL}
\eeq
with the intercepts given by $\, \alpha_p = 1 + \epsilon \,$ and
$\, \alpha_{R} = 1 - \eta \,$. The simplicity of \ref{DL} has made the DL model
fit of total cross sections a standard reference for models of total, elastic
and diffractive cross sections \cite{five}.

Although the DL model fares reasonably
well when fitting to  $\,pp\,$ and $\,p\bar{p}\,$
total cross sections,
its $\chi^2$/d.o.f. becomes considerably larger that that of other models
\cite{Kang} when fitting both the total cross sections and the real
parts of the scattering amplitudes.
This lead two of us (JRC and KK) with S. K. Kim to propose a slight
generalisation of the DL model,
lifting the degeneracy of the reggeon trajectories\cite{six}:
\be
\sigma_{tot} (s) = X s^{\epsilon} + Y_+ s^{-\eta_{+}} \pm Y_- s^{-\eta_{-}}
\label{CKK}\ee
The last two terms
represent the exchanges of non-degenerate $\, C=+1 (a,f)\, ,$ and $\, C= -1
(\rho , \omega )\, $ meson trajectories, with intercepts $\,
\alpha_p=1-\epsilon
$ and $\, \alpha_{\pm}=1-\eta_{\pm}$ respectively. The sign of the $\, Y_-\,$
term flips when fitting $\,p \bar{p}\,$ data compared to $\,pp\,$ data. The
real parts of the forward elastic
amplitudes are calculated from analyticity\cite{KN}.
We shall refer to this model as $CKK$.

Despite its successes, the above parametrisation \cite{six} left some questions
unanswered. First of all, one had to introduce a filtering of the
data to produce a reasonable $\chi^2$/d.o.f., irrespectively of the model. Such
a filtering inevitably leads to a bias, which we could not evaluate precisely.
The present analysis uses an expanded and revised dataset  \cite{dataset}
which does not call for a selection of datapoints.
Secondly, the analysis in \cite{six} was limited to $pp$ and $\bar pp$ data,
and could not consider questions related to factorisability and universality
of trajectories.

We present here the preliminary results
of a joint fit to $pp$, $\bar pp$, $\pi^\pm p$, $K^\pm p$, $\gamma p$ and
$\gamma\gamma$ cross sections and hadronic $\, \rho$-parameters. The latter two
processes are particularly important in the study of DIS events at HERA.
We will see that the soft-pomeron intercept obtained in \cite{six} is
reproduced by global fits to all available soft data from these reactions.
Also, in order to see if the soft data are enough to establish the
existence of a simple Regge pole for the pomeron, we present also fits to a
typical analytic amplitude model \cite{AS},
i.e., Model A2 out of many possible parameterizations in \cite{Kang},
\be
\sigma_{tot} (s) = \Lambda \left(A + B \ln ^2 (s/s_0)\right)
+ Y_+ s^{-\eta_{+}} \pm
Y_- s^{-\eta_{-}}
\label{RRL2}
\ee
which we shall call $RRL2$ in the following, and where the last two terms
represent the lower Regge trajectory terms of $\, C = \pm 1 \,$ as before.
We will see that the $RRL2$ fit is indistinguishable from the $CKK$ simple-pole
one. 
Thus while one cannot conclude that the pomeron is a
simple pole, the claims concerning eventual problems with unitarity
\cite{Levin} is not supported either.
\section{Results of the simple-pole fit: the $CKK$ model}
We first give the results for the simple-pole fit (model $CKK$). As before \cite{six} we fit the data above an energy cut-off $\sqrt{s_{min}}$ and require
that the results be stable w.r.t. variations in that cut-off, and that
the $\chi^2$/d.o.f. be of order 1. Furthermore, in order to get a (slightly)
better fit at low energy, we use the variable $\tilde s\equiv (s-u)/2$ in
eqs.~(\ref{CKK}, \ref{RRL2}). 

We assume the intercepts $\epsilon$, $\eta_+$, and $\eta_-$ to be universal,
and the couplings are then related through charge conjugation by:
 $X_{h_1h_2}=X_{h_1\bar h_2}$,  $Y_\pm^{h_1h_2}=\pm Y_\pm^{h_1\bar
h_2}$.
Hence $Y_-^{h\gamma}=0$.
We shall rewrite the pomeron couplings in the following forms, which
make their properties more transparent:
\beq
X_{pp}&=&x_{pp}\times{3\over 2} X_{\pi p}, \ \ \ \ \ \ 
X_{Kp}=x_{Kp}\times X_{\pi p}, \label{xb}\\
X_{\gamma p}&=&x_{\gamma p}\times {g_{elm}}^2\left[{1\over f_\rho^2}
+{1\over f_\omega^2}+{1\over f_\phi^2}\right](1+\delta) X_{\pi p}
\approx
 x_{\gamma p} \times {X_{\pi p}\over 213.9}, \label{xc}\\
X_{\gamma\gamma}&=&x_{\gamma\gamma}\times {X_{\gamma p}^2\over X_{pp}}.
\label{xd}
\eeq
The parameters $x_{h_1h_2}$ are expected to be of order 1, because (\ref{xb})
reflect the additive quark counting, (\ref{xc}) comes from generalised
vector-meson dominance (GVMD) \cite{GVMD},
where the contribution of off-diagonal
terms $\delta$ is expected to be about 15\%,
and (\ref{xd}) is a prediction from the factorisation property of the pomeron
couplings.

The number of data points is shown in Fig.~1(a), and the resulting
$\chi^2$/d.o.f. in Fig.~1(b), upon changing $\sqrt{ s_{min}}$ from
3 to 20 GeV.
\begin{figure}
\vglue 1cm
\hglue 3.5cm (a)\hglue 8.5cm (b)\\
\vglue -1.5cm
\centerline{
\psfig{figure=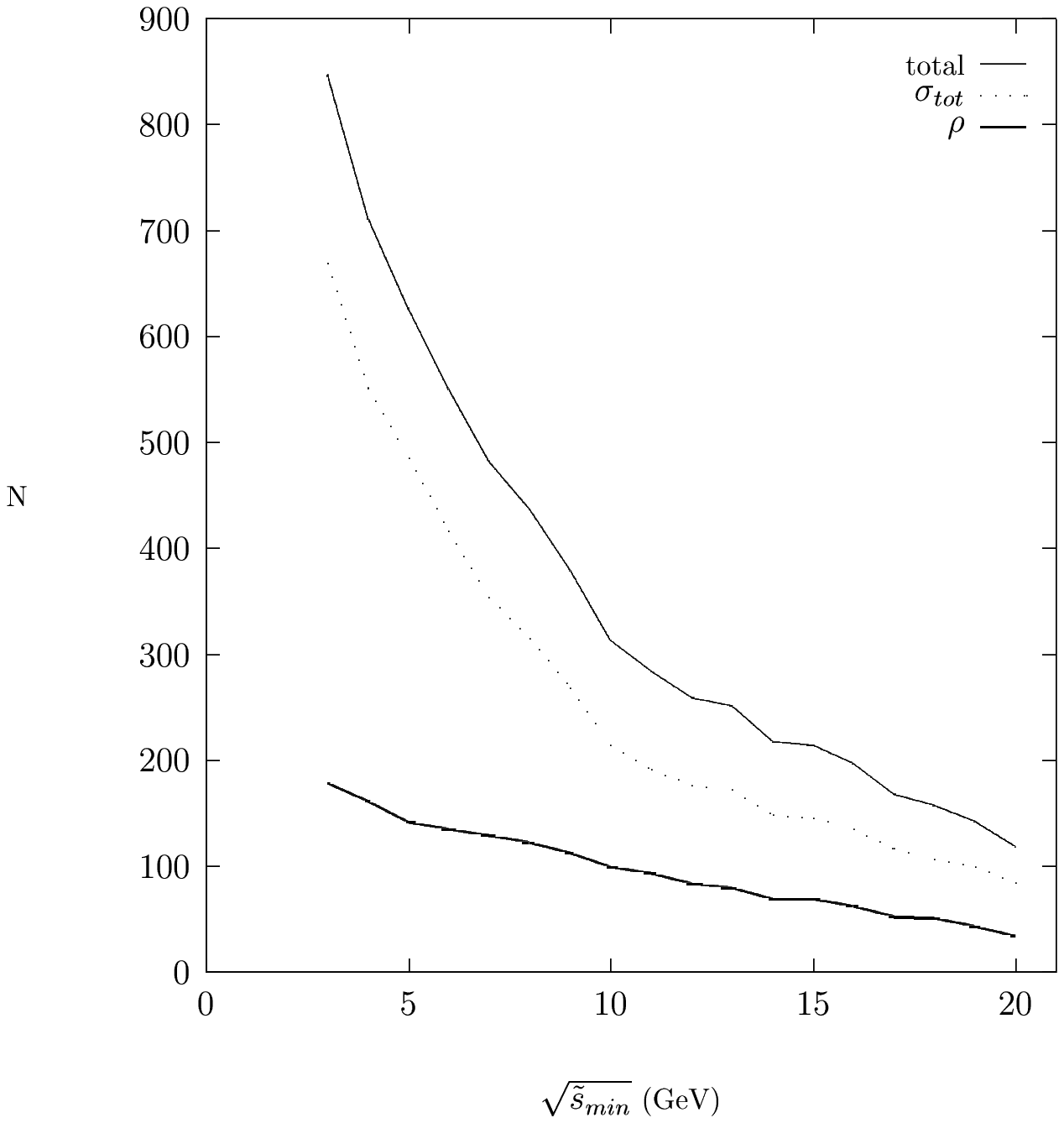,height=7cm}\hglue 1cm
\psfig{figure=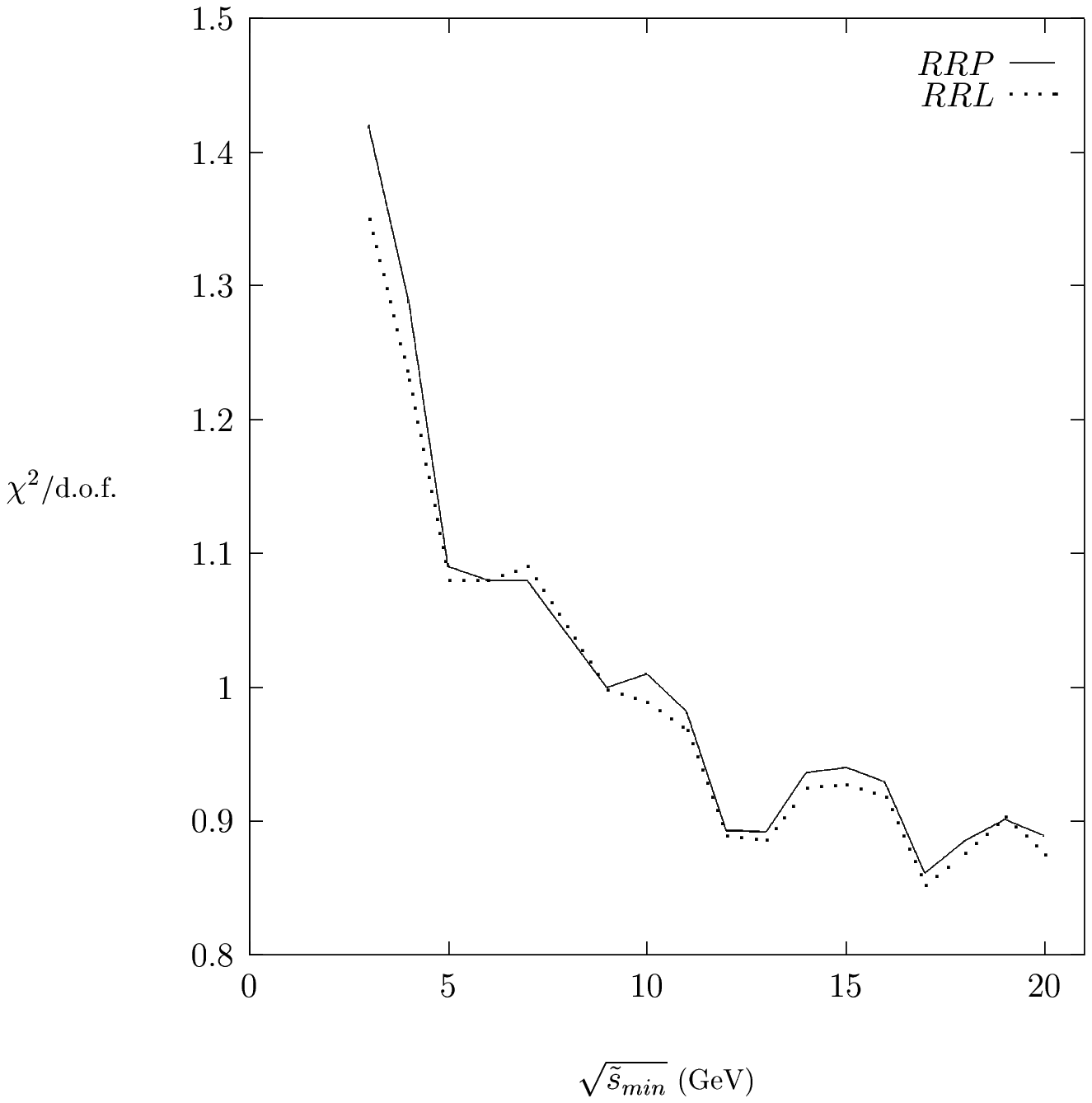,height=7cm}
}
\begin{quote}
Figure 1:  The number of points included in the fit (a) and the resulting
$\chi^2$/d.o.f. as a function of the minimum energy $\sqrt{ s_{min}}$.
\end{quote}
\end{figure}
Clearly, the fit is bad for small energies. This is expected, as there is
no reason to neglect the effects of lower trajectories and thresholds then. We
also see
that values of 1 or smaller for the $\chi^2$/d.o.f. can be achieved
for $\sqrt{ s_{min}} \geq 9$ GeV.

The second criterion concerns the stability of the
parameters. We show in Figure 2 the intercepts of the three trajectories
entering
(\ref{CKK}).
\begin{figure}
\vglue 1cm
\hglue 4.5cm (a)\hglue 6.5cm (b)\\
\vglue -1.5cm
\centerline{
\psfig{figure=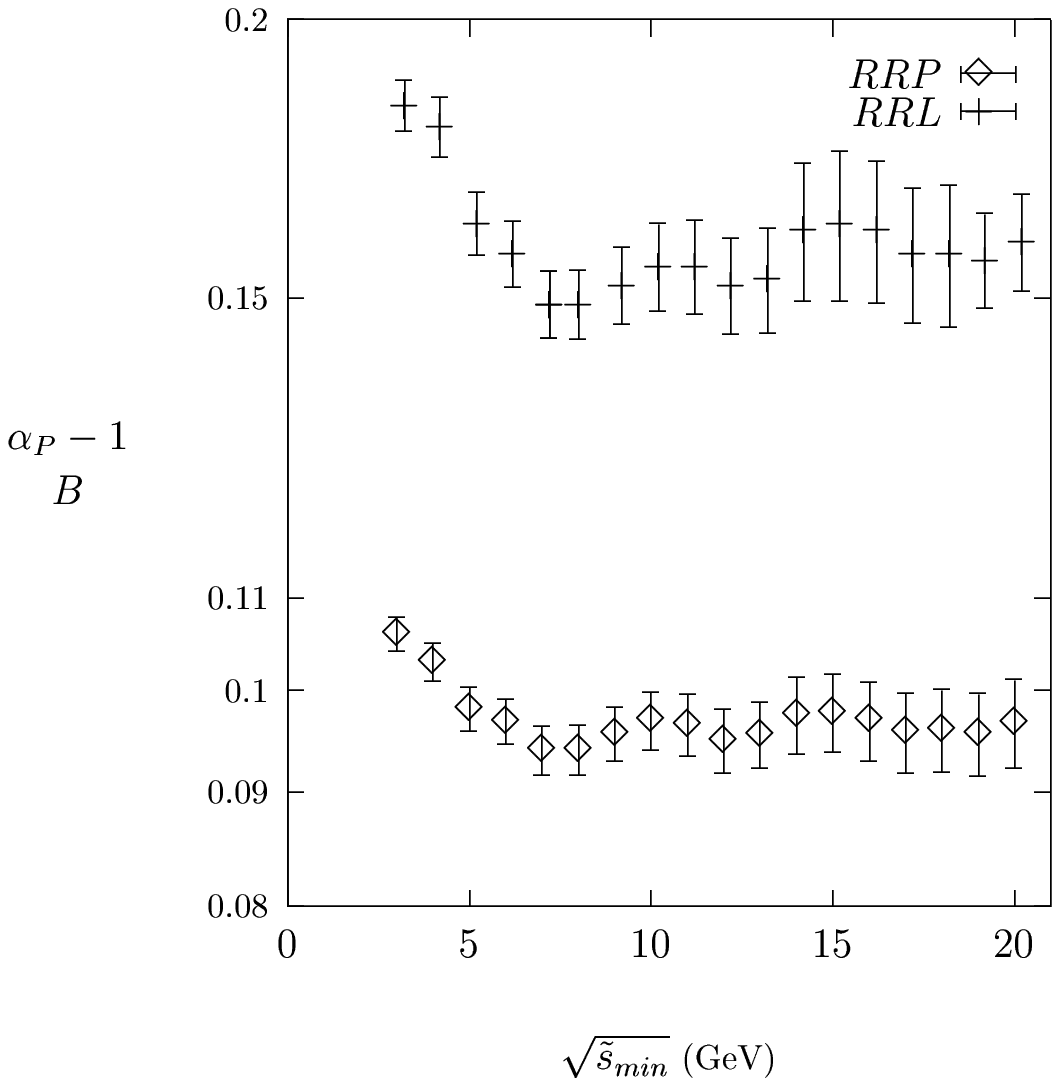,height=7cm}\hglue 1cm
\psfig{figure=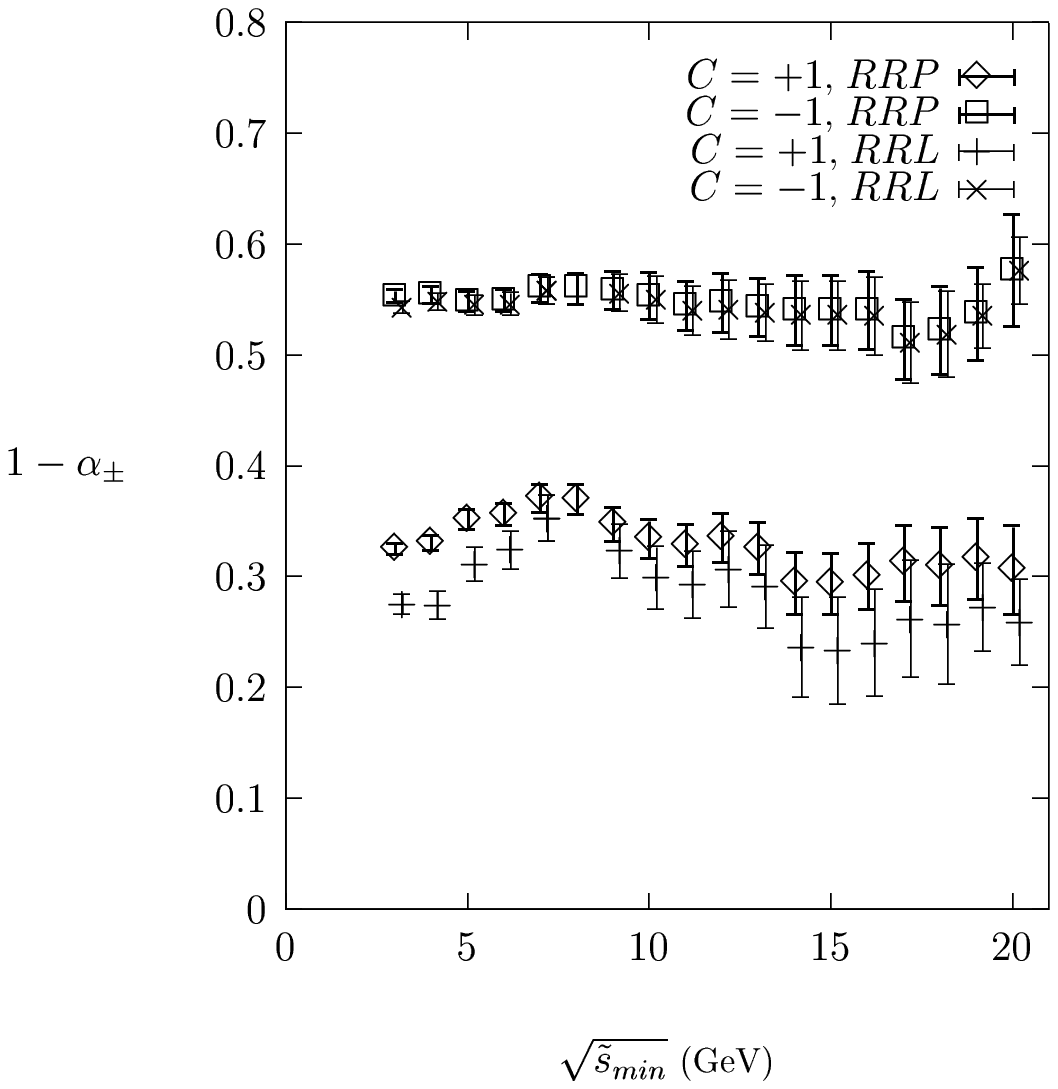,height=7cm}
}
\begin{quote}
Figure 2:  The value of the pomeron intercept ($RRP$) and of the coefficient of
the $\log^2 s$ ($RRL$) (a) and of the intercepts
of the $a/f$ ($C=+1$) and $\rho/\omega$ ($C=-1$) trajectories.
\end{quote}
\end{figure}
One sees that these parameters are stable once the energy is above 9 GeV.
Also one obtains a larger pomeron intercept
for a smaller energy cut-off.
The intercepts from the global fit to all soft data are the same
as the results of \cite{six} above 9 GeV which is based on $\, pp\,$ and
$\,p\bar{p}\,$ data alone, and this
justifies to some extent  the statistical data treatment
and numerical procedure
employed in \cite{six}.

With the increased dataset from all available reactions, the
errors of the parameters can be narrowed, compared to those of
\cite{six} where the errors correspond to a change
of 5 units in $\chi^2$.
As in \cite{six}, we need both $\, C=\pm 1\,$ meson trajectories,
which are non-degenerate, primarily because of the
constraints coming from fitting the $\, \rho \,$ parameters.

We show in Fig.~3 how the value of the pomeron coupling $X_{\pi p}$
together with those of $x_{h_1h_2}$ depends on the minimum energy.
\begin{figure}
\vglue 1cm
\hglue 3.cm (a)\hglue 8cm (b)\\
\vglue -1.5cm
\centerline{
\psfig{figure=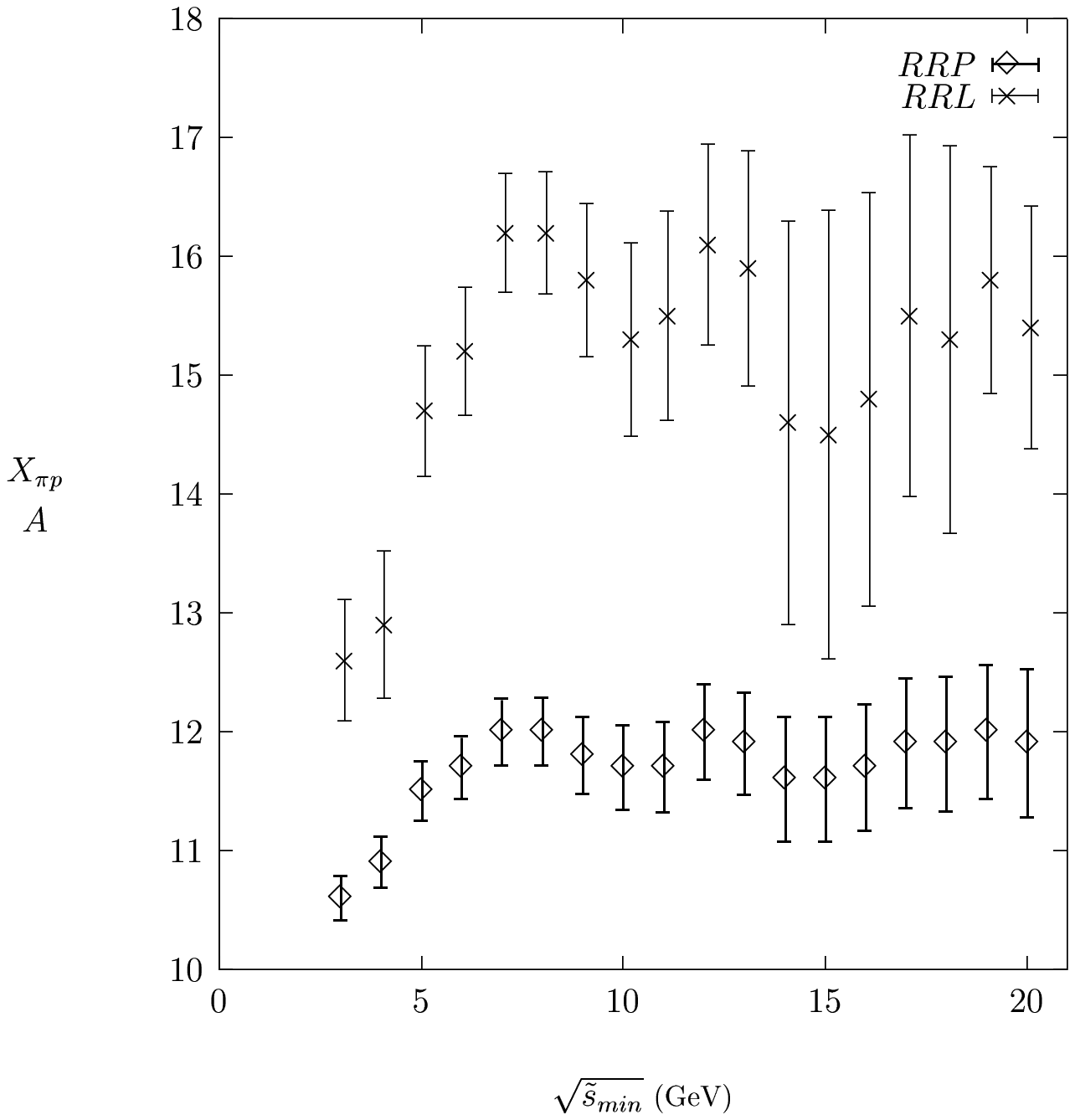,height=7cm}\hglue 1cm
\psfig{figure=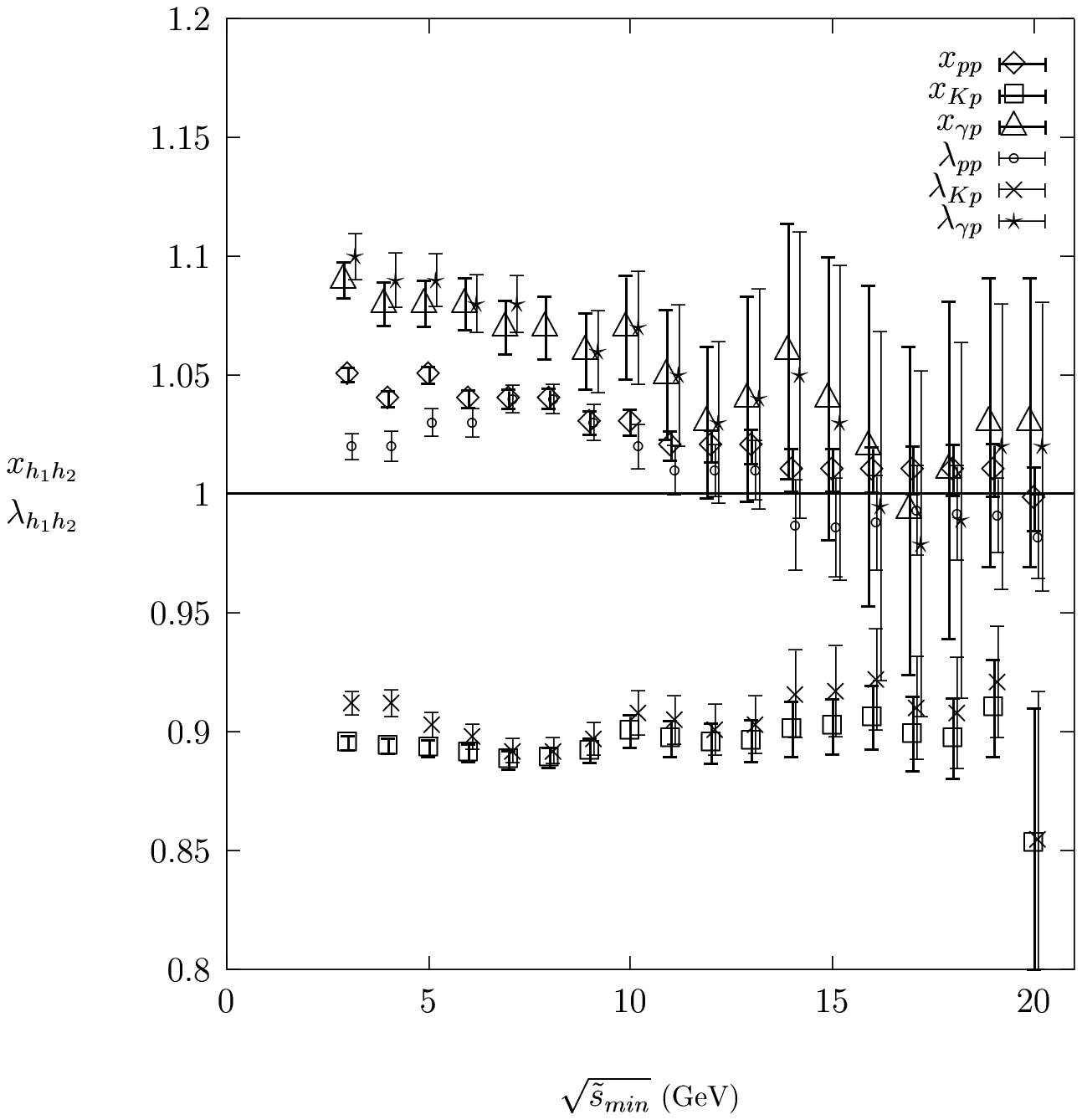,height=7cm}
}
\begin{quote}
Figure 3:  The value of the pomeron $\pi p$ coupling ($RRP$) and of the
constant term $A$ ($RRL$)  (a) and of the $x$ and $\lambda$
couplings defined in Eqs.~(\ref{xb} - \ref{xd}) and (\ref{lambda}).
\end{quote}
\end{figure}

Note that the pomeron couplings become stable also for energies greater
than 9 GeV. As the intercepts and pomeron couplings are the most important
parameters we give our best fit results in Table 1 for $\sqrt{ s}\geq 9$ GeV,
where we have most statistics. If we set
\beq
X_{pp,p\bar{p}} &=&  9 \beta_{qp}^2 , \,
\quad X_{ \pi^{\pm} p} = 6\beta_{qp}^2 ,
\quad X_{K^{\pm} p} = (\beta_{sp} + \beta_{qp}) (3\beta_{qp}) \nonumber\\
X_{\gamma p} &=&  3\beta_{qp}\beta_{\gamma p},
\quad X (\gamma\gamma) = \beta_{\gamma p}^2
\eeq
where $\, \beta_{qp}\,$
denotes the pomeron coupling to any of the non-strange quarks $\, u\,$ and $\,
d \,$,  $\, \beta_{sp}\,$  the pomeron-strange quark coupling, and
$\,\beta_{\gamma p}\,$ the
pomeron-photon coupling, we get numerically from the global fit result
of Table 1 that $\, \beta_{qp} = 1.39(2) , \beta_{sp} = 1.12(2)\,$ and $\,
\beta_{\gamma p} = 0.0139(3)$.
We see from this and also from Fig.~ 3(b) that the pomeron couplings respect
factorizability based on the additive quark counting within a few percents, and
that the pomeron
coupling to the $s$ quark is 15\% lower than that to $u$ and $d$. Also GVMD
works well.

However, it is worth pointing out that the couplings of the lower trajectories
are not as stable as those of the pomeron.
The quark counting and factorisation are violated by about
50\% for $C= +1$ Reggeon couplings, although GVMD still works well. The quark
counting is totally off for the $C=-1$ Reggeon couplings.
This problem of instability is easy to
understand:
the Regge couplings are basically representing the low energy nature of
 the data where they have to compete with the secondary Regge contributions and
multi-Regge correction terms. At high-energies they compete with the pomeron,
which determines most of the cross section. Therefore there seems to be
no best cut-off for their determination. Only a model including more
trajectories (and many more parameters!) and elastic and inelastic threshold
effects might achieve the stability.
The couplings in Table 1
 are those determined with this 9 GeV energy cutoff where the $\chi^2$, the
pomeron
parameters and Regge intercepts
 are showing stability. In this respect, the error determination based
on a $\, \chi^2 \,$ variation of  one unit may be underestimating the true
errors, and certainly is in the case of the couplings of the lower
trajectories.
\begin{figure}
\begin{center}
{\footnotesize
\begin{tabular}{|c|c|c|c|c|c|}   \hline
 $\epsilon$         &  $\eta_+ $     &  $\eta_-$ & $\chi^2$/d.o.f.&
statistics&\\ \hline\hline
$ 0.096\pm  0.003$   & $0.35\pm 0.02$ & $0.56\pm 0.02$ &  1.00&268 &\\
\hline\hline
    & $pp$              & $\pi p$            & $Kp$           & $\gamma p$
   & $\gamma\gamma$\\ \hline
$X$ (mb)   & $ 18.3\pm 0.6$ & $11.8\pm 0.3$  & $10.5\pm 0.3$  & $0.059\pm
0.002$ & $(1.5\pm 0.2)\times 10^{-4}$\\
$Y_+ $ (mb)& $61\pm 2$      & $26\pm 1$      & $14\pm 1$      & $0.12\pm 0.01$
& $(4 \pm 3) \times 10^{-4}$\\
$Y_-$ (mb) & $36\pm 3$      & $8\pm 1$       & $14\pm 1$ & &\\
\hline\hline
process&$\chi^2/N$, $\sigma_{tot}$ (N)&$\chi^2/N$ ,$\rho$ (N)&process
&$\chi^2/N$, $\sigma_{tot}$ (N)&$\chi^2/N$, $\rho$ (N)\\\hline
$pp$         &  1.02 (74)&    1.25  (59)&$K^+ p$      & 0.534 (22)&0.635 (7) \\
$\bar pp$    &  1.16 (33)&   0.504 (11)&$K^- p$      &  0.824 (28)&1.97  (5) \\
$\pi^+ p$    & 0.551 (24)&    2.17 (7)&$\gamma p$   &  0.627 (25)&\\
$\pi^- p$    &  1.13 (47)&  0.939 (23)&$\gamma\gamma$&   0.265 (15)&\\
\hline\hline
\end{tabular}}
\end{center}
\begin{quote}
Table 1: the values of the parameters of the hadronic amplitude in model
$CKK$ (\ref{CKK}), corresponding to a cut off $\sqrt{ s}\geq9$ GeV,
and the values of the individual $\chi^2$ of the various processes, together
with the number of points $N$.
\end{quote}
\end{figure}
We also show the $\chi^2$ per data points, and the number of data points,
for each process fitted to. One can see that, as in our previous work
\cite{six}, the
$\chi^2$ is a little high for some of the sub-processes.
We have shown
in \cite{six} that this has nothing to do with the model, but rather with
the dispersion of the data. Filtering the data for these two processes did
not change the determination of the parameters.
We shall demonstrate in another way
in the next section that this is probably
due to inconsistencies within the data, and
that this does not affect our conclusions.
\begin{figure}
\vglue 1cm
\hglue 5.5cm (a)\hglue 5.5cm (b)\\
\vglue -1.5cm
\centerline{
\psfig{figure=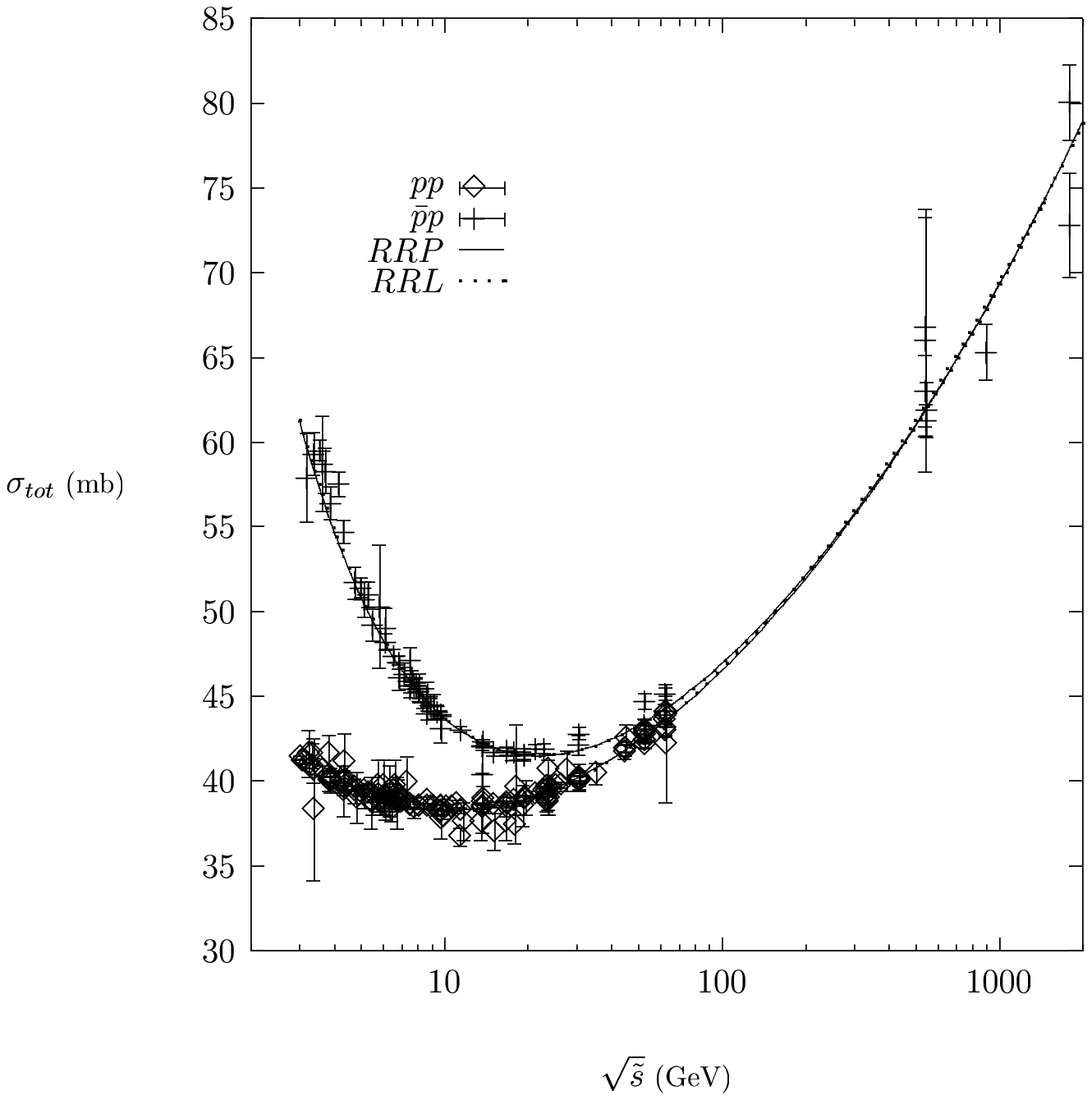,height=7cm}\hglue 1cm
\psfig{figure=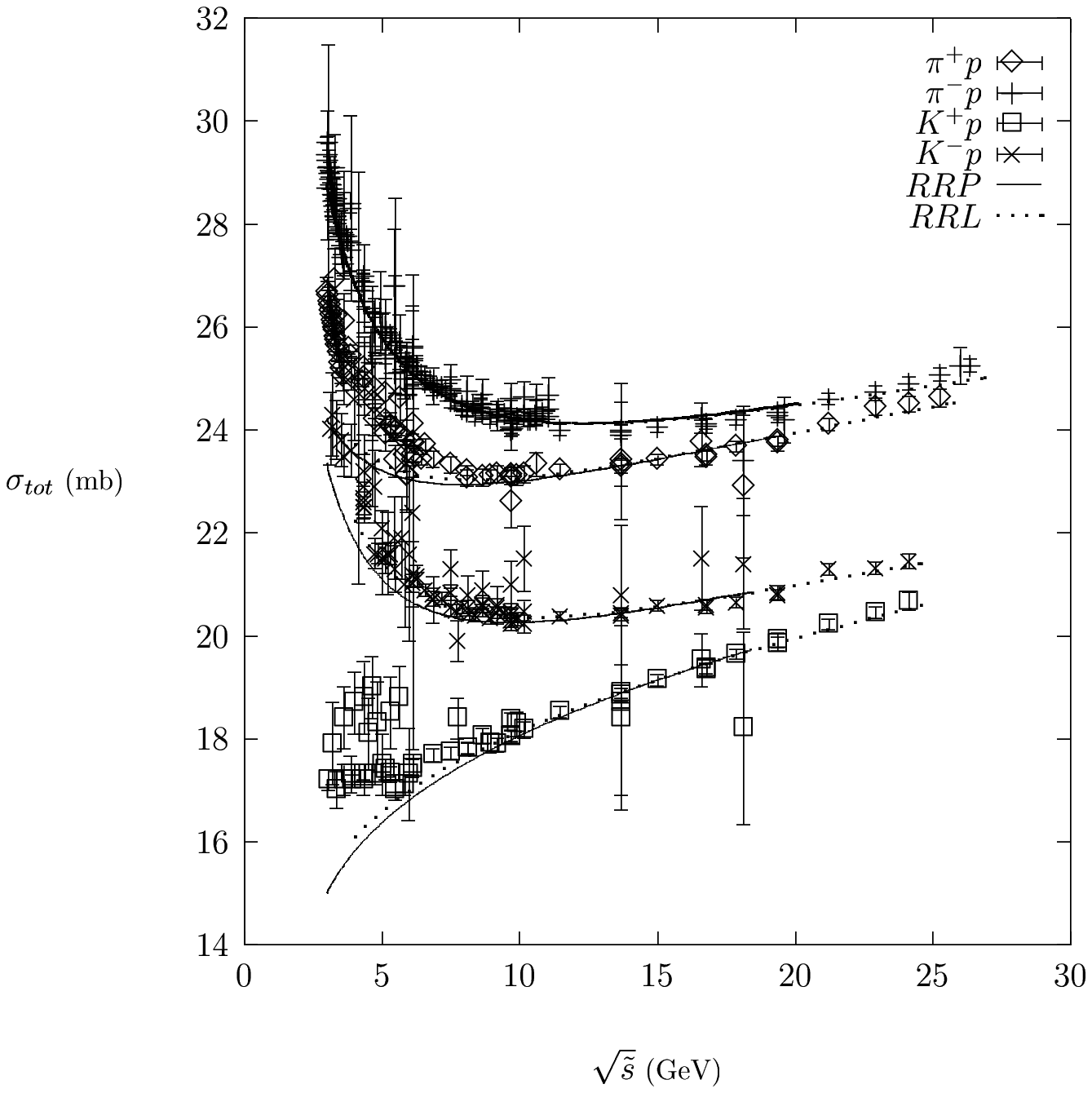,height=7cm}
}
\vglue 1cm
\hglue 2.5cm (c)\hglue 9.0cm (d)\\
\vglue -1.5cm
\centerline{
\psfig{figure=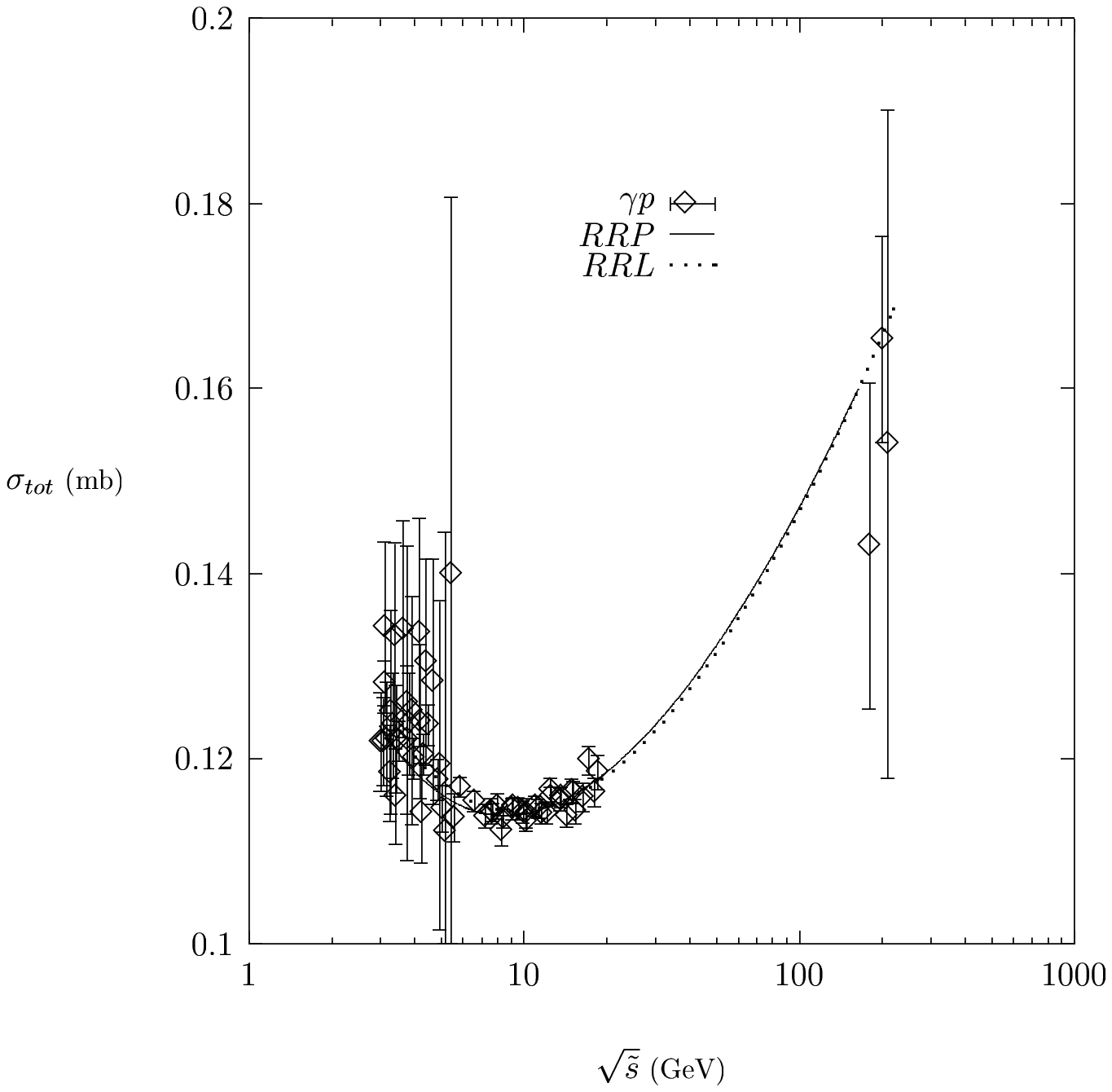,height=7cm}\hglue 1cm
\psfig{figure=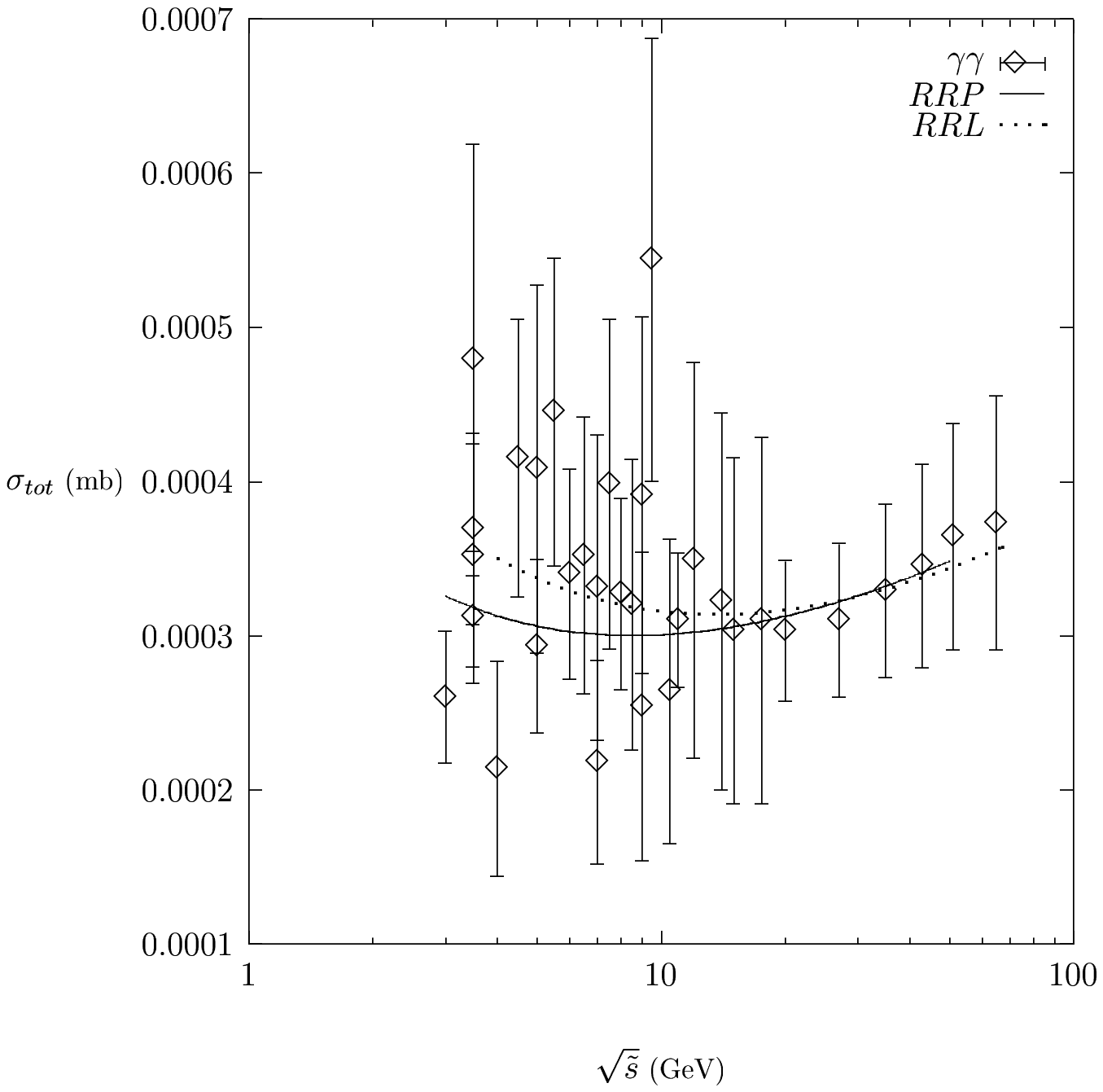,height=7cm}
}
\begin{quote}
Figure 4:  The total cross sections corresponding to the parameters of Table 1.
The fits have been performed for $\sqrt{ s}\geq 9$ GeV.
\end{quote}
\end{figure}
\begin{figure}
\centerline{
\psfig{figure=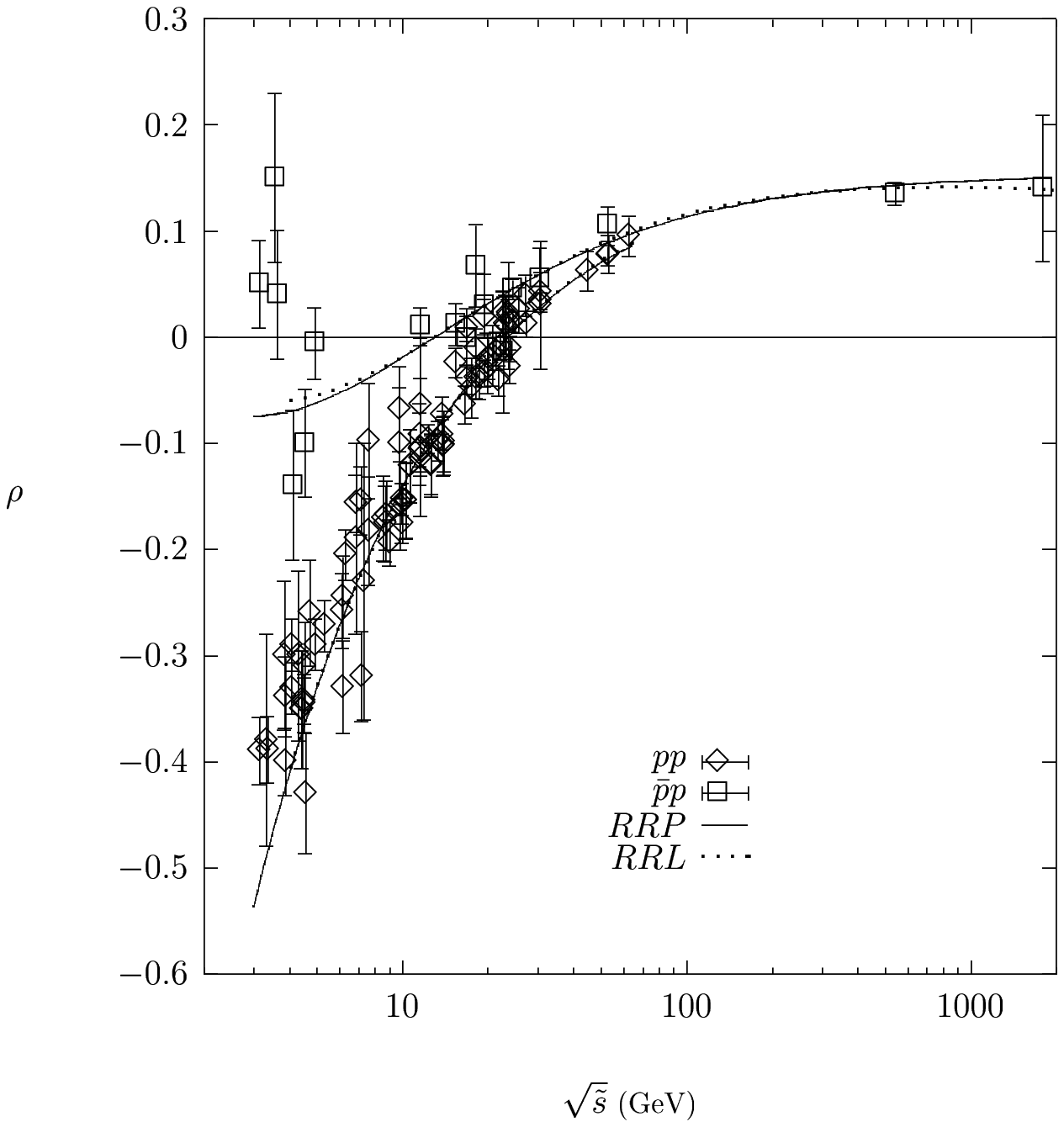,height=7cm}\hglue 1cm
\psfig{figure=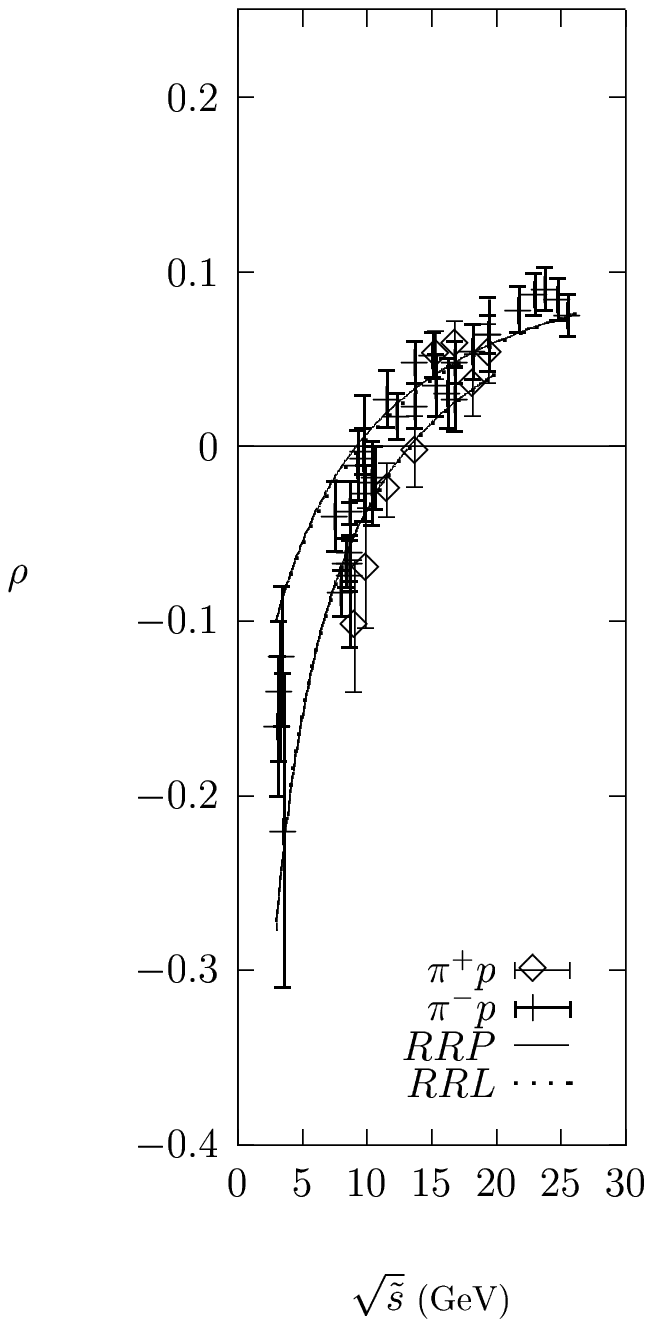,height=7cm}
\psfig{figure=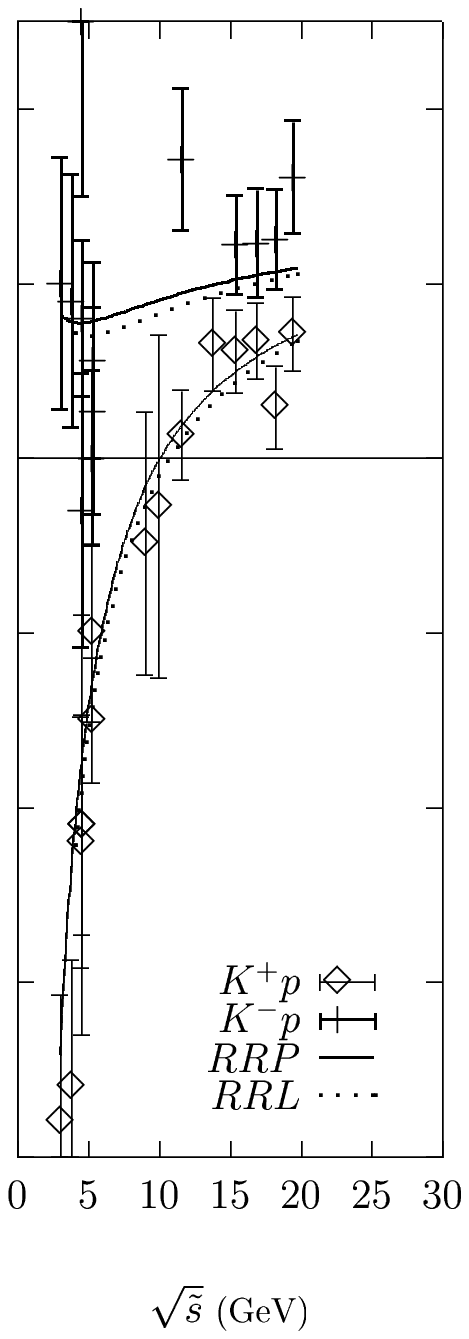,height=7.1cm}
}
\begin{quote}
Figure 5:  The $\rho$ parameters corresponding to the parameters of Table 1.
The fits have been performed for $\sqrt{ s}\geq 9$ GeV.
\end{quote}
\end{figure}
The fits for
the total cross sections and $\rho$-parameters respectively for $\,
\sqrt{ s_{min}}\geq 9$
GeV, and extrapolated to  $\, \sqrt{ s_{min}}\geq 3$
GeV.  are shown in Figs.~4 and 5.
Although the value of $\, \chi^2$/d.o.f. is bad in the low-energy region
(it goes above 2),
and thus is statistically unacceptable, the fits look deceptively satisfactory.
\section{Comparison with the $RRL2$ model}
While the Regge pole hypothesis works surprisingly well with the soft data up
to
the Tevatron energy, this is not the only parameterization that is successful
in term of $\, \chi^2$/d.o.f. :
there have been a number of successful analytic amplitude representations at
the phenomenological level \cite{Kang} based on analyticity and with the
asymptotic behavior $\ln^2 s$ or $\ln s $ for the total cross sections, when
appropriately modified by the meson trajectory contributions as in (\ref{CKK}),
which
could give equally good or better fits to the $\, pp\,$ and $\, p\bar{p}\,$
data as shown in \cite{Kang}.  In these models, one may also regard the
asymptotic $\,
\ln^2 s$ or $\, \ln s\,$ terms as an effectively unitarised form \cite{FFKT} of
the bare
pomeron term of
$\,s^\epsilon$.
We know that a simple pole will eventually violate the Froissart-Martin
bound.
We show here
first that in the region of available data, the two descriptions of Eq.~(\ref{CKK})
and (\ref{RRL2})
are indistinguishable. Furthermore,
the instability
present in the lower trajectory couplings of the simple-pole fit has nothing to
do with the assumptions regarding the pomeron.

We present here the results of the fit to all soft data of
total cross sections and $\rho$-parameter for the modified Amaldi-Schubert
($RRL2$) model (\ref{RRL2}) suitably factored to exhibit
the factorization property by the $\, \ln^2 s\,$ term.
In (\ref{RRL2}), $\, \Lambda \,$ will be different for different reactions and
will
follow the factorization property based on the additive quark counting, if the
$\,\ln^2 s$ term is to represent the pomeron exchange contributions.  Here,
the parameters $\, A\,$ and $\, B\,$ and the meson trajectory intercept
parameters $\, \eta_+$ and $\,\eta_- \,$ are taken to be universal for all
reactions, while  the factorization parameter $\, \Lambda\,$ which will be set
to 1
for $\,\pi^\pm p\,$.
In analogy with (\ref{xb} - \ref{xd}) we define:
\beq
\Lambda_{pp} = \lambda_{pp}\times{3\over 2} ~ , ~ \Lambda_{Kp} = \lambda_{Kp},
\nonumber\\
\Lambda_{\gamma p}
\approx {\lambda_{\gamma p} \over 213.9} ~ , ~ \Lambda_{\gamma\gamma} =
\lambda_{\gamma\gamma}\times {\Lambda_{\gamma
p}^2\over \Lambda_{pp}}
\label{lambda}\eeq
In order to simplify our discussion, and in order to have the same number of
parameters for both fits, we set\footnote{It is possible to get slightly
better fits below $s = 9$ GeV$^2$ if one lets this parameter free, but
it reaches unphysical values of the order of 100 MeV$^2$ or smaller, and
the stability of the fit is not improved.} $s_0=1$ GeV$^2$. We again use
$\tilde s$ instead of $s$ in Eq.~(\ref{RRL2}) to improve the fit at small energy.

We proceed as for the simple-pole fit. We see from Figs. (1-5) that identical
problems and successes are present in this case. It is interesting to observe
that
the couplings of the $a/f$ trajectory go down a little, but
remain unstable, whereas the $C=-1$ contribution remains identical. One amazing
outcome is
that despite this variation, the two fits are identical. As shown in Fig.~1,
the $\chi^2$/d.o.f. are the same. If we again settle on the parameters
corresponding to $\sqrt s_{min} = 9$ GeV, shown in Table 2, we obtain the
dashed
curves of Figs. 4 and 5, which are almost identical to those of the simple-pole
fit. This has two important consequences. First, the simple-pole assumption
is one of the possibilities, but not the only one. One has to realise however
that the property of quark counting and factorisation, exhibited by both
fits,
is hardly understandable outside of a simple-pole ansatz. Hence it is
the physics that must make us prefer this fit and the possibility to extend it
to elastic and diffractive events, and not simply the quality of reproduction
of the $t=0$ data. Furthermore, as already mentioned, the indistinguishability
of the simple pole from the $\log^2 s$ proves that effectively there cannot be
any problem with unitarisation. 
As the difference in total cross sections at the LHC energies is at most 6 mb
it seems unlikely that such an effect will be detectable, even if we assume
that
the total cross section will be reliably measurable within the approved CERN
program.

\begin{figure}
\begin{center}
{\footnotesize
\begin{tabular}{|c|c|c|c|c|c|}   \hline\hline
$A$ (mb)& $B$ (mb)       &  $\eta_+ $     &  $\eta_-$ & $\chi^2$/d.o.f. &\\
\hline\hline
$15.8 \pm 0.3 $   &  $0.15\pm 0.06$& $0.32\pm 0.02$ & $0.56\pm 0.02$ &  1.00
&\\ \hline\hline
             & $pp$           & $\pi p$            & $Kp$           & $\gamma
p$         & $\gamma\gamma$\\ \hline
$\Lambda$    & $ 1.55\pm 0.01$ & $1$               & $0.90\pm 0.07$  & $4.96\pm
0.08)\times 10^{-3}$ & $(1.3\pm 0.2)\times 10^{-5}$\\
$Y_+ $ (mb)& $51\pm 2$        & $20\pm 1$          & $10\pm 1$      & $0.088\pm
0.008$ & $(3 \pm 3) \times 10^{-4}$\\
$Y_-$ (mb) & $36\pm 3$        & $7\pm 1$           & $14\pm 1$ & &\\
\hline\hline
\end{tabular}}
\end{center}
\begin{quote}
Table 2: the values of the parameters of the hadronic amplitude in model
$RRL2$ (\ref{RRL2}), corresponding to a cut off $\sqrt{ s}\geq 9$ GeV.
and the values of the individual $\chi^2$ of the various processes. The number
of points, the statistics and the $\chi^2$ for each process
are the same as for Table~1.
\end{quote}
\end{figure}
\section{Concluding remarks}
There is no room in any of the considered models for another
$C=+1$ Regge trajectory with intercept $0.8\leq \alpha_{new}\leq 1.2$.
All couplings are then set to a very small value. Note however that this
conclusion is possible only after inclusion of the real parts in
the dataset. Total cross sections only allow (and slightly favour) an
extra flat contribution.

As for the presence of a hard pomeron with intercept $\approx 1.4$
\cite{twopoms},
there was no room in $pp$ and $\bar p p$ data for such an object \cite{six},
and
we obtain a ($1\sigma$) upper bound on its
coupling: $X_{new}/X\leq 5\times 10^{-9}$. However, such a large intercept
will certainly call for a strong unitarisation.
It is puzzling that in fact the presence of such a term is favoured both by the
$\pi p$ and $Kp$ data, particularly for the measured $\rho$ parameters.
Notice from Table 1 that the individual
$\chi^2$ were rather high for the fits to the $\rho$ parameters of  $\pi$ and
$K$. These $\chi^2$ per data point get lowered to from 2.17 (resp. 1.97) to
1.51 (resp. 1.42) in the presence of a hard pomeron. The global
fit then gets a $\chi^2$/d.o.f. of 0.89 instead of 1.00. The hard
pomeron intercept is fitted to $0.38\pm 0.13$, with a coupling of the order of
1.5\% that of the soft pomeron. The $\gamma p$ data do not favour such a term,
and the errors on the $\gamma\gamma$ data are large enough to allow for it.
But as the total $\gamma p$ and $\gamma\gamma$ data are under reconsideration,
and as
unitarity effect may set in early, it is hard to draw any conclusions about
these.

The data does not allow further $C=-1$ trajectories. The quality of the fit is
not improved by the introduction of one such trajectory, and all couplings are
less that 1 per thousandth of the soft-pomeron coupling. We have also tried
to stabilize the couplings of lower trajectories through the introduction
of new $C=\pm 1$ trajectories with lower intercepts. However, the fit depends
then on
26 parameters, and the data is not constraining enough to draw any
firm conclusion.

Concerning the existing conflicting data, first of all, the long-standing
problem
of the measurement of the total cross section at the Tevatron \cite{Tevatron}
cannot be resolved by this method. As can be seen from Figure 4, the fit
chooses the middle points of the two measurements. Because of our choice of
$\chi^2$, which does not privilege higher-energy data, this conclusion is
stable independently of removal of
either the CDF or the E710 point from our dataset.
Secondly, there is a well-known uncertainty about the HERA total cross
section\cite{HERA}.
Our fit favours the H1 measurement.  Imposing GVMD exactly leads to exactly the
same conclusion. This is expected in view of the ZEUS analysis of low-$Q^2$
$F_2$ data \cite{lowq}.
Finally, as there may be some problem regarding the value of the $\gamma\gamma$
cross sections \cite{LEP}, we can remove the
high-energy $\gamma\gamma$ data from our dataset,
and we can predict the $\gamma\gamma$ cross section
 imposing factorisation $x_{\gamma\gamma}=y_{\gamma\gamma}=1$. This leads to
the dashed curve of Fig.~6.
\begin{figure}
\centerline{
\psfig{figure=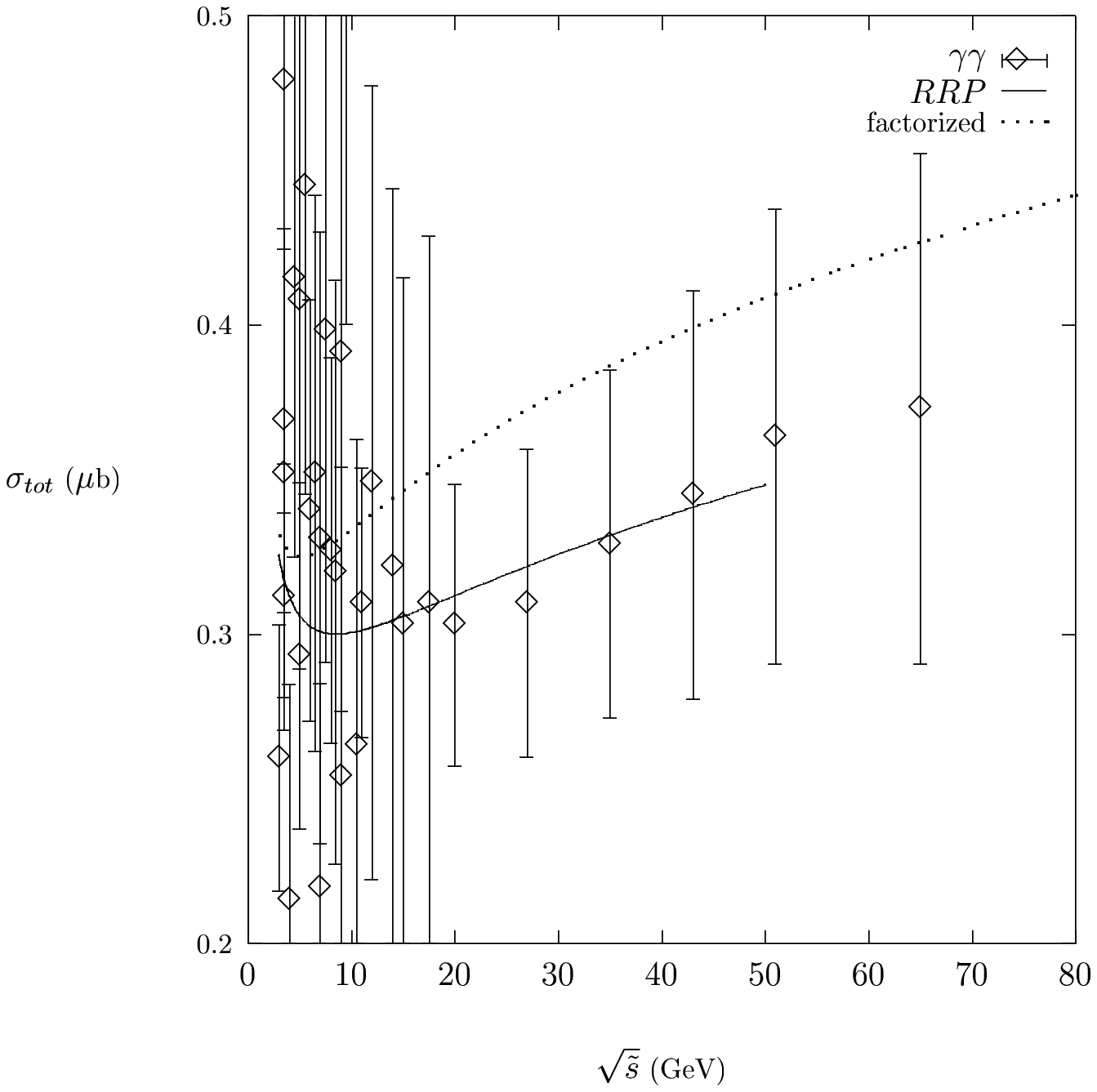,height=7cm}\hglue 1cm
}
\begin{quote}
Figure 6:  The $\gamma\gamma$ total cross section at LEP,
assuming factorisation.
\end{quote}
\end{figure}
We see that the factorisation hypothesis favours higher numbers than the
published L3 data
but compatible with the preliminary OPAL measurement\footnote{New L3 $183 GeV$
and OPAL measurements are now consistent as reported to ICHEP'98 in Vancouver
\cite{OPAL}. We would like to thank A. De Roeck for bringing this to our
attention.}

We have shown that the soft pomeron produces very good fits to $t=0$ data,
once the energy is bigger than 9 GeV.
From our new compilation of data points, and from the 264 points above 9 GeV,
we determined the pomeron intercept to be $1.096\pm 0.03$, in agreement
with the conclusions of \cite{six}.
Lower $C=\pm 1$ trajectories are non-degenerate, and have intercepts
given in Table 1. The determination of these parameters is stable and
reliable, as is that of the pomeron couplings,
but the interplay between $C=+1$ contributions makes
the determination of the couplings of the $a/f$ trajectories problematic.
Finally, $t=0$ data are not sufficient to rule out
other models of forward scattering amplitudes, but the factorisation
and quark counting properties which seem to be well respected are difficult
to be understood outside of the context of simple poles. Further details of the
work along with the fits to other analytic amplitude models and the results of
the efforts to ameliorate the instability of the lower Reggeon couplings will be reported elsewhere\cite{CKC}.

One of us ({\bf KK}) would like to acknowledge Professor R. Vinh Mau and his group for warm hospitality extended to him at LPTPE, Universit{\'e} P. \& M. Curie (Paris6) during a part of the 1998 summer where a part of this work was carried out.


\begin{thebibliography}{99}
\bibitem{DL}A. Donnachie and P.V. Landshoff, {\it Phys.Lett.} {\bf B296} (1992)
227.
\bibitem{Collins} P.D.B. Collins, \underline{An Introduction
to Regge Theory and High Energy
Physics} (Cambridge University Press, Cambridge: 1977).
\bibitem{PDG}Review of Particle Physics, Particle Data Group (C. Caso et al.)
{\it The European Physical Journal} {\bf C3} (1998) 1.
\bibitem{one}ZEUS Collaboration (J. Breitweg et al.) {\it Eur. Phys. J.} {\bf
C2} (1998) 247-267; (M. Derrick et al.) {\it Phys. Lett.} {\bf B315} (1993)
481;
 {\it Phys. Lett.} {\bf B338}(1994) 483;  {\it Phys. Lett.} {\bf B356} (1995)
129; {\it Z. Phys.} {C68} (1995) 569;\\
H1 Collaboration (C. Adloff et al.) {\it Z. Phys.} {\bf C76} (1997) 613-629;
(C. Adloff et al.) {\it Z. Phys.} {\bf C76} (1997) 613.
\bibitem{two}Samim Erhan and Peter E. Schlein, e-Print Archive: hep-ph/9804257
(March 1998), submitted to Phys. Lett. B; UA8 Collaboration (R. Bonino et al.)
{\it Phys. Lett.} {\bf 211} (1988) 239; (A. Brandt et al.) {\it Phys. Lett.}
{\bf B297} (1992) 417;
\bibitem{twopoms}A. Donnachie and  P.V. Landshoff, e-Print Archive:
hep-ph/9806344.
\bibitem{glueballs} WA102 Collaboration (D. Barberis et al.) e-Print Archive:
hep-ex/9805018, WA91 Collaboration (S. Abatzis et al.) {\it Phys. Lett.} {\bf
B324} (1994) 509.
\bibitem{DoLa}A. Donnachie and  P.V. Landshoff, {\it Nucl. Phys.} {\bf B267}
(1986) 690, {\it Nucl. Phys.} {\bf B244} (1984) 322,
P.D.B. Collins, F.D. Gault and A. Martin {\it Nucl. Phys.} {\bf B85} (1975)
141.
\bibitem{four}A. Donnachie and P.V. Landshoff, {\it Nucl. Phys.} {\bf B231}
(1984) 189;  {\it Nucl. Phys.} {\bf B267} (1986) 657; {\it Phys. Lett.} {\bf
B296} (1992) 227.
\bibitem{five}Review of Particle Physics, Particle Data Group (R.M. Barnett
et al.), {\it Phys. Rev.} {\bf D54} (1996) 1.
\bibitem{Kang} See for instance K. Kang,  P. Valin, and A.R. White, {\it Nuovo
Cim.} {\bf 107A} (1994) 2103; K. Kang and S. K. Kim, in Proc. VIth Blois
Workshop, \underline{Frontiers in Strong Interactions} (Chateau de Blois,
France, June 1995), Brown-HET-1008 and hep-ph/9510438.
\bibitem{six} J.R. Cudell, K. Kang and S.K. Kim, {\it Phys. Lett.} {\bf B395}
(1997) 311.
\bibitem{KN}K. Kang and B. Nicolescu, {\it Phys. Rev.} {\bf
D11} (1975) 2461.
\bibitem{dataset}Data used were extracted from the PPDS accessible at
http://pdg.lbl.gov or
http://wwwppds.ihep.su:8001/ppds.html.
Computer readable data files are also available at
http://pdg.lbl.gov.
\bibitem{AS}Amaldi et al., {\it Phys.Lett.} {\bf B66} (1977) 390.
\bibitem{BKW} M.M. Block, K. Kang, and A.R. White, {\it Mod. Phys.} {\bf A7}
(1992) 4449.
\bibitem{Levin}E. Gotsman, E.M. Levin, and U. Maor, {\it Phys.Rev.} {\bf D49}
(1994) 4321.
\bibitem{GVMD}G. Shaw, {\it Phys. Rev.} {\bf D47} (1993) 3676, {\it Phys.
Lett.} {\bf B228} (1989) 125,
P. Ditsas and G. Shaw, {\it Nucl. Phys.} {\bf 113} (1976) 246, J.J. Sakurai and
D. Schildknecht,
{\it Phys. Lett.} {\bf B40} (1972) 121.
\bibitem{FFKT} J. Finkelstein, H.M. Fried, K. Kang, and C.-I Tan, {\it Phys.
Lett.} {\bf B232} (1989) 257.
\bibitem{Tevatron}E710 collaboration (N. Amos et al.), {\it Phys. Lett.} {\bf
B243} (1990) 158.
\bibitem{HERA}H1 Collaboration (T. Ahmed et al.)  {\it Phys. Lett.} {\bf B299}
(1993) 374; ZEUS Collaboration
(M. Derrick et al.), {\it Z. Phys.} {\bf C63} (1994) 391,
{\it Phys. Lett.} {\bf B293} (1992) 465.
\bibitem{lowq} B. Surrow, Ph. D. dissertation, DESY-THESIS-1998-004 (January
1998).
\bibitem{LEP}L3 Collaboration (M. Acciarri et al.) {\it Phys. Lett.} {\bf B408}
(1997) 450,
OPAL collaboration, preliminary results, physics note 320 (Sept. 1997),
http://www.cern.ch/Opal/pubs/pn/html/pn320.html.
CDF collaboration (F. Abe et al.) {\it Phys. Rev.} {\bf D50} (1991) 5550.
\bibitem{OPAL} S. S{\"o}ldner-Rebold, Proc. ICHEP'98, Vancouver, Canada, July
22-30, 1998, hep-ex/9810011; A. De Roeck, Proc. 4th Workshop on Small-x and
Diffractive Physics, Fermilab, 17-20 September 1998.
\bibitem{CKC}J.R. Cudell, V.V. Ezhela, K. Kang, S.B. Lugovsky, and N.P. Tkachenko, to be published.
\end{thebibliography}
\end{document}